\documentclass[
    ,final            
  ]
  {aipproc}

\layoutstyle{6x9}

\input pubboard/babarsym
\RequirePackage{xspace}

\newcommand{\acp}{\ensuremath{\calA_{ch}}}
\newcommand{\calB}{\ensuremath{{\cal B}}}

\newcommand{\DE}{\ensuremath{\Delta E}}

\newcommand\etal{{\it et al.}}

\newcommand{\bfig}{\begin{figure}[htbpc!]}
\newcommand{\efig}{\end{figure}}
\newcommand\bef{\begin{figure}}
\newcommand\edf{\end{figure}}

\newcommand\beq{\begin{equation}}
\newcommand\eeq{\end{equation}}
\newcommand\bear{\begin{array}}
\newcommand\enar{\end{array}}
\newcommand\beqa{\begin{eqnarray}}
\newcommand\eeqa{\end{eqnarray}}
\newcommand\ben{\begin{enumerate}}
\newcommand\een{\end{enumerate}}

\newcommand{\UfourS}{\ensuremath{\Upsilon(4S)}}

\newcommand{\Kst}{\ensuremath{K^*}}

\newcommand{\Kstz}{\ensuremath{\Kstarz}}

   \newcommand{\rhoz}{\ensuremath{\rho^0}}

\newcommand{\fetaK}{\ensuremath{\eta K}}
\newcommand{\etaK}{\ensuremath{\B\ra\fetaK}}

\newcommand{\fetaKst}{\ensuremath{\eta K^{*}}}
\newcommand{\etaKst}{\ensuremath{\B\ra\fetaKst}}

\newcommand{\fetarhop}{\ensuremath{\eta\rho^+}}

\newcommand{\fetapK}{\ensuremath{\etapr K}}
\newcommand{\etapK}{\ensuremath{\B\ra\fetapK}}

\newcommand{\fetapKz}{\ensuremath{\etapr K^0}}
\newcommand{\etapKz}{\ensuremath{\Bz\ra\fetapKz}}

\newcommand{\etapKst}{\ensuremath{\B\ra\etapr K^{*}}}

\newcommand{\fetaprhop}{\ensuremath{\etapr\rho^+}}

\usepackage{color}
\usepackage{graphpap}
\newcommand{\plotdir}{plots}
\newcommand{\beginpixoverlay}[3][]{ 
  \\[-#3pt]
  \begin{picture}(#2,#3)(0,0)
  \def\noOpt{}\def\testit{#1}\ifx\testit\noOpt{}%
  \else%
  \graphpaper(0,0)(#2,#3)
  \fi
}

\begin{document}

\begin{flushleft}
\babar-PROC-05/055 \\
\end{flushleft}

\title{Rare B Decays and B Decay Dynamics}


\author{William T. Ford}{
  address={Department of Physics \\University of Colorado\\ Boulder, CO 80309-0390}
}


\begin{abstract}
I present some recent measurements of $B$ meson decay rates to leptonic and
charmless hadronic final states, as well as of \CP-violation charge
asymmetries and other features.  I sketch the theoretical frameworks
used to predict these, and indicate the level of agreement of the
estimates with experiment.
\end{abstract}

\maketitle


\section{Introduction}

The dominant channels through which ground-state hadrons containing $b$
quarks decay are those in which the flavor change is $b\ra c$.  The more
rare channels are suppressed by the smaller value of the CKM matrix
element $V_{ub}$ relative to $V_{cb}$, or by dynamical effects such as
off-shell partons in loops, helicity conservation, the OZI mechanism,
higher-order electroweak coupling factors, etc.

\section{Leptonic decays:  $B\ra\tau\nu_\tau$}

For the pure leptonic process $\Bp\ra\ell^+\nu_\ell$ the branching
fraction is given by the familiar formula 
\beq
\calB(\Bp\ra\ell^+\nu_\ell) = \frac{G_F^2m_B}{8\pi}m_\ell^2
\left(1-\frac{m_\ell^2}{m_B^2}\right)^2f_B^2\left|V_{ub}\right|^2\tau_B.
\label{eq:BFBtaunu}
\eeq
Here $G_F$ is the Fermi constant, $f_B$ is the $B$ decay constant,
$\tau_B$ the lifetime, and $V_{ub}$ the $b\ra u$ CKM matrix element.
The factor $m_\ell^2$ reflects the suppression by helicity
conservation.  Because of this effect the most sensitive measurements
are made with
$\ell=\tau$, for which the branching fraction expected in the Standard
Model is around $10^{-4}$.

Recent new measurements by Belle \cite{blBtaunu} and BaBar
\cite{bbBtaunu} provide new limits on the
branching fraction.  These are done by reconstructing the recoil $B$
meson in one of its more common decays, to hadrons in the case of Belle
and (mostly) to semileptonic decays by BaBar.  
In the reconstruction of semileptonic decays the neutrino four-momentum
can be determined up to an ambiguity from the $B$ flight direction,
which however is a small effect because of the small $B$ meson momentum in
the \UfourS\ frame. 
With one $B$ meson accounted for,
candidates are selected for each of the main $\tau$ decays.  Even though
the result does not give the full energy and momentum of this $B$,
because one or two neutrinos are missing, this construction exhausts all
charged tracks and neutral energy clusters in the event.  One examines
the residual calorimeter energy for a low-end delta function smeared by
imperfect photon reconstruction, in contrast with a broad tail from
background.

Neither experiment finds a significant signal, Belle observing 39 events
against an expected background of 31.4, and BaBar observing 150 events
where 130.9 background are expected.  The 90\% C.L. upper limits are
\begin{center}
\begin{tabular}{lcl}
\hline
	& $N(\BB)/10^6$	& $\calB(\Bp\ra\tau^+\nu_\tau)$	\\
\hline
Belle	& 232			& $<1.8\times10^{-4}$	\\
BaBar	& 275			& $<2.8\times10^{-4}$ ($1.28^{+1.15}_{-1.08}$)\\
\hline
\end{tabular}
\end{center}
BaBar quote a limit $f_B<0.34\ \gev$, to be compared with a lattice
calculation \cite{LQCDfB} $f_B=0.20\pm0.03\ \gev$.  So the experiments
are approaching the sensitivity needed to detect a significant signal.

\begin{figure}
  \label{fig:ckmFitBtaunu}
  \includegraphics[width=.45\textwidth,bb=0 -20 521 539]{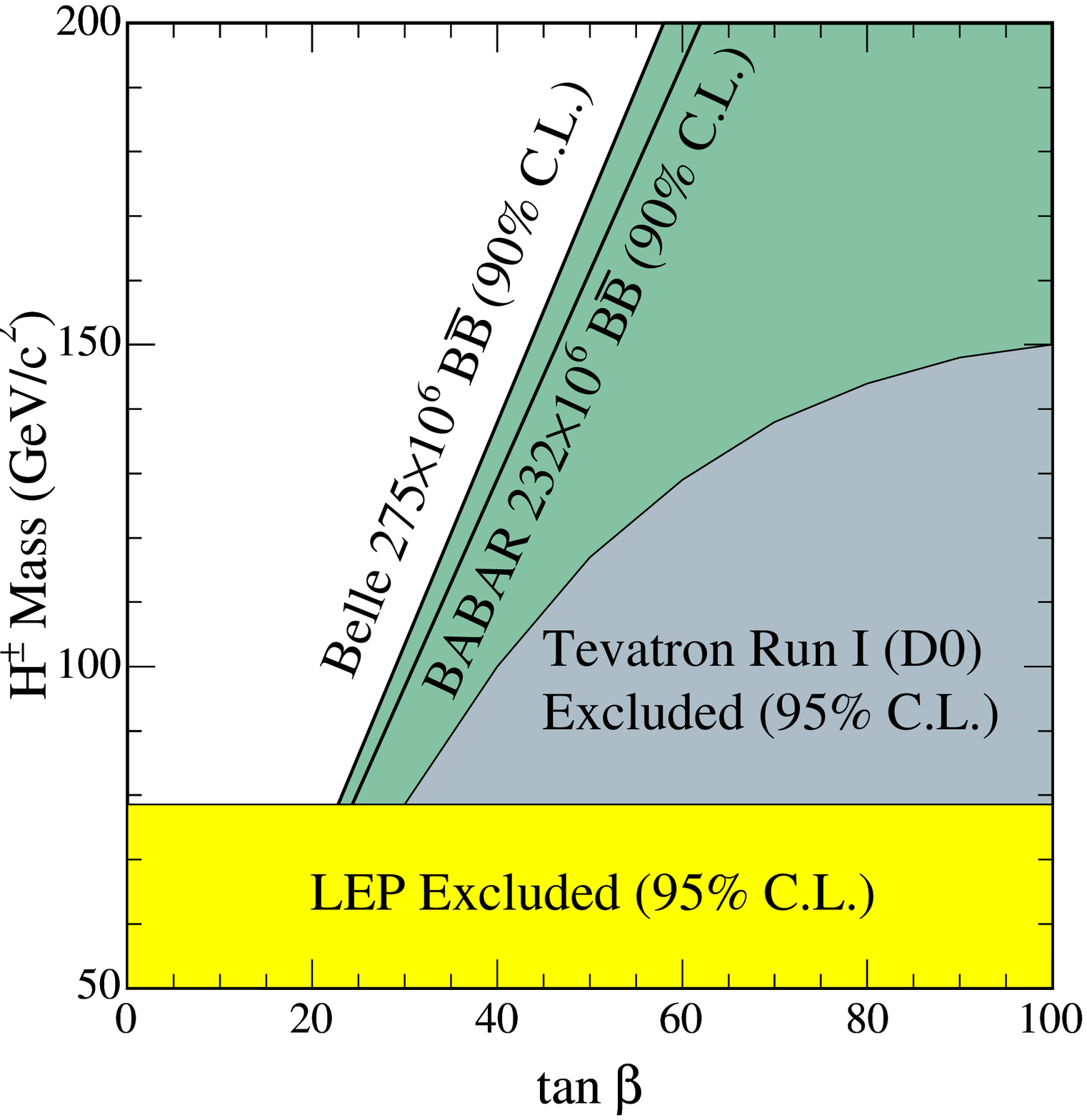}
  \includegraphics[width=.55\textwidth]{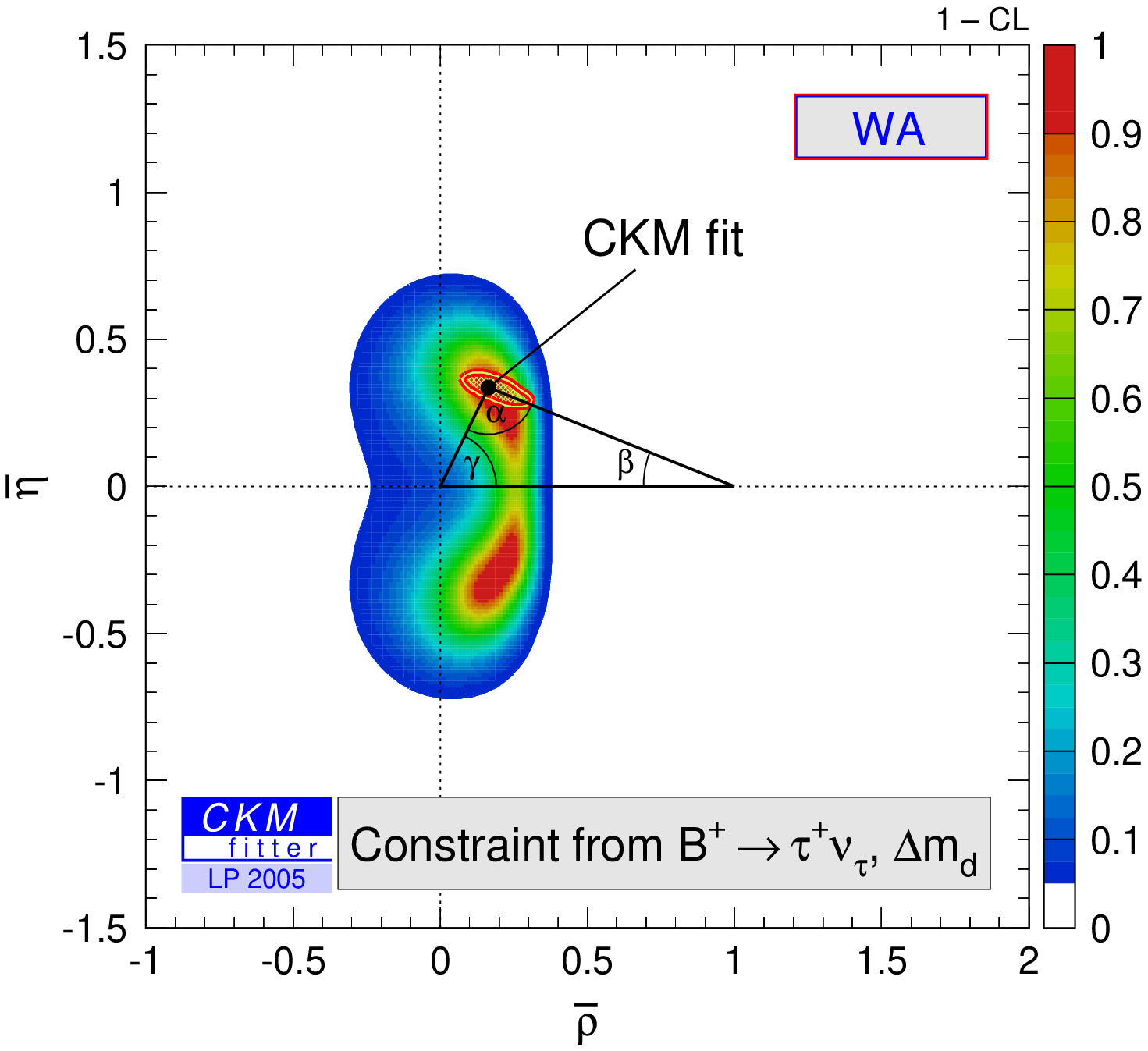}
  \caption{Limits on charged Higgs mass and ratio of VEVs (left), and
constraint on $(\bar{\rho},\bar{\eta})$ from
$\calB(\Bp\ra\tau^+\nu_\tau)$ combined with $\Delta
m(B_d)$ (right).} 
\end{figure}

The ratio of this branching fraction to the mass difference $\Delta
m(B_d)$ is proportional to
$\left|V_{ub}\right|^2/\left|V_{tb}\right|^2$, the coefficient having
only linear dependence on the ``bag parameter'' $B_b$, and none on
$f_B$.  The resulting constraint on the CKM Wolfenstein parameters
$(\bar{\rho},\bar{\eta})$ is shown in Fig.\ \ref{fig:ckmFitBtaunu}
(right-hand plot) \cite{bib:ckmFitBtaunu}.  Alternatively, fixing the
CKM matrix we deduce a limit on the contribution of a possible charged
Higgs to this decay, in the minimal two-Higgs doublet model.  The left
plot of Fig.\ \ref{fig:ckmFitBtaunu}, from \cite{blBtaunu}, shows the
limits in the space of 
Higgs mass versus the ratio $\tan{\beta}$ of vacuum expectation values
(VEVs) of the charged and neutral doublets.

\section{Charmless hadronic B decays}

Since semileptonic and radiative decays generally bear on the subject of
the magnitudes of CKM matrix elements covered in the talk
by Kinoshita, I turn now to hadronic $B$ meson decays, concentrating
on the ``rare'' charmless channels.  This is an area of vigorous
activity to map out the many channels experimentally and to understand
them theoretically, or at least to characterize them
phenomenologically.  The presence of non-perturbative hadronic effects
complicates the picture, but the large energy release in these heavy to
light decays provides the possibility to control those uncertainties.

At the parton level these processes are mediated by amplitudes
represented by diagrams like those shown in Fig. \ref{fig:diags}.
\begin{figure}
  \label{fig:diags}
  \includegraphics[width=0.5\textwidth,angle=270,origin=cl,bb=91 63 521
701,clip]{\plotdir/diags.eps}
  \caption{Representative Feynman diagrams for charmless $B$ meson
decays:  (a) external tree ($\Delta S=0,\ T$); (b) color-suppressed
tree ($|\Delta S|=1,\ C^\prime$); (c) gluonic penguin ($P^\prime$); (d)
flavor singlet penguin ($S^\prime$).}  
\end{figure}
One phenomenological approach to the estimation of decay rates and
charge asymmetries identifies a reduced matrix element with each of the
parton topologies and relates their contributions to the various decay
modes via flavor-SU(3) symmetry \cite{suprunPP,suprunVP}.  Seven
independent reduced diagrams and the CKM angle $\gamma$ are fit to all
of the available data within each of the final state particle classes
$P$--$P$, $P$--$V_s$, $P_s$--$V$, where $P$ is a pseudoscalar meson, $V$
is vector meson, and the subscript indicates the meson containing the
spectator quark.  The seven topologies are the four named in the
Fig. \ref{fig:diags}\ caption, plus weak annihilation ($a$), $W$-boson
exchange ($e$) and penguin annihilation ($pa$).  This picture is found
to be quite compatible with the data as indicated by the fit
chisquares, and Figures \ref{fig:Kpi}-\ref{fig:dS0}.

The direct calculation of decay rates and charge asymmetries begins with
the effective Hamiltonian written as an operator product expansion
(OPE) \cite{bblRMP}.  For a $b\ra s$ transition: 
$$
{\cal H}_{\rm eff} =
\frac{G_F}{\sqrt{2}}\sum_{p=u,c}V^*_{ps}V_{pb}\left(
 C_1Q^p_1 + C_2Q^p_2 + \sum_{i=3}^{10}C_iQ_i + C_{7\gamma}Q_{7\gamma}
 + C_{8g}Q_{8g}\right) + {\rm h.c.}.
$$
The operators correspond to terms in the full theory at parton level as
\begin{itemize}
 \item $Q^p_{1,2}$: current-current operators from $W$ exchange
 \item $Q_{3\dots 6}$: local 4-quark QCD penguin operators
 \item $Q_{7\dots 10}$: local 4-quark electroweak $\gamma, g, Z$
penguin, and $W$ box operators
 \item $Q_{7\gamma}$: electromagnetic dipole operator
 \item $Q_{8g}$: chromomagnetic dipole operator,
\end{itemize}
while $C_i,C_{7\gamma},C_{8g}$ are the Wilson coefficient functions.
The factorization of each term facilitates the calculation by separating
factors calculable, to next-to-leading order (NLO) in the strong coupling
constant $\alpha_s$, from QCD and the renormalization group.  This
separation is however scale- and scheme-dependent, requiring that the
matrix elements be calculated to matching order in the same scheme and
scale.  The matrix elements include the problematic long-distance
effects.  

The factorization ansatz for dealing with the hadronic matrix elements
employs the concept of ``color transparency'': because of the large
$Q$-value in a heavy quark decay the daughter mesons fly from the region
of their formation so quickly that their soft hadronic interactions are
suppressed (by a factor of order $\Lambda_\mathrm{QCD}/m_b$).  The
matrix element becomes a product of a form factor representing the
transition of the $B$ to one meson and a decay constant representing the
creation from vacuum of the other daughter meson.  Some of the earlier
applications \cite{ali} treat quark masses and the effective number of
colors as free parameters in fits to data.

\begin{figure}
  \label{fig:bn_theorem}
  \includegraphics[width=0.6\textwidth]{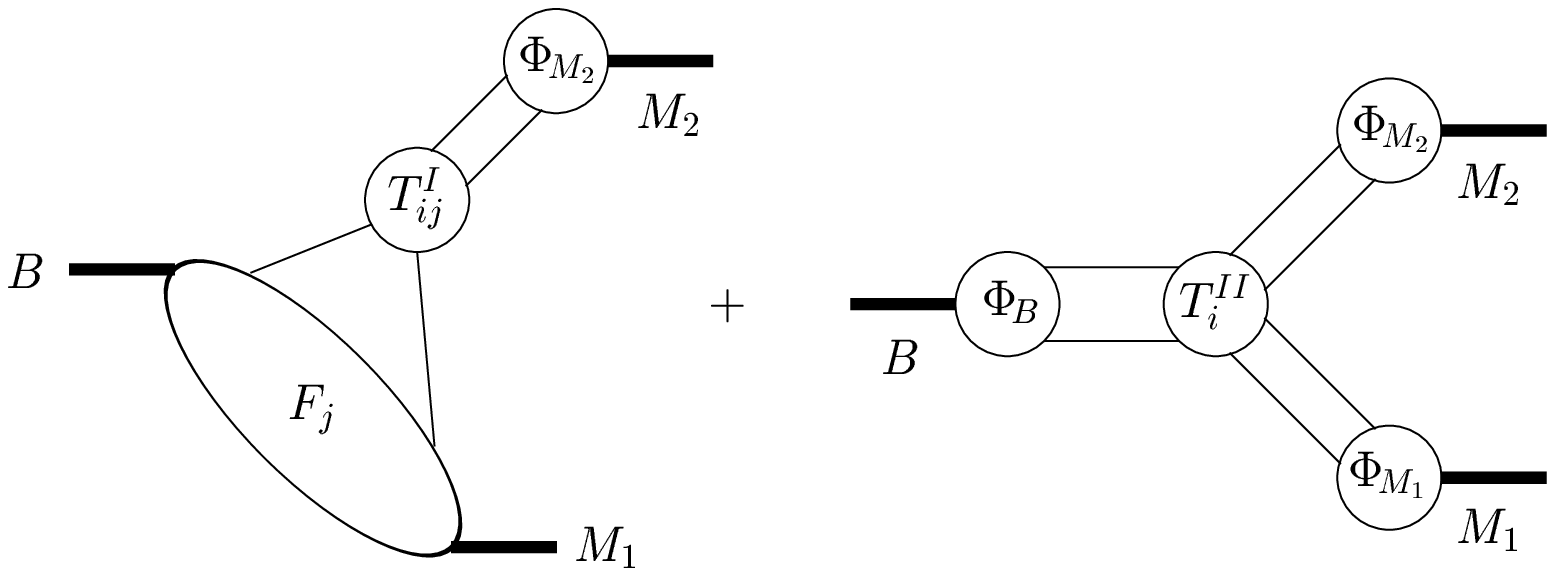}
  \caption{QCD factorization formula, from \cite{beneke}.}
\end{figure}

The naive factorization method has been improved upon (``QCD
factorization'', QCDF \cite{BBNS,beneke}) with a more general formula shown
schematically in Fig. \ref{fig:bn_theorem}.
The factors $T^{I}$ and $T^{II}$ are the hard-scattering kernels,
calculated in the heavy quark limit at NLO.  The $B$ to
daughter meson form factor $F$, and light-cone parton distribution
functions $\Phi$ are inputs to the calculation.  Non-perturbative
effects are absorbed into these factors.  A copy of the first
term with $M_1\leftrightarrow M_2$ is implied.  The second term accounts
for interactions with the spectator quark.  

An alternative improvement on naive factorization is provided by the
``perturbative QCD'' framework (pQCD) \cite{pQCD,keum,kouSanda}.  In
this approach the 
treatment of the parton transverse momentum serves to control endpoint
singularities in the parton distribution functions, allowing the
calculation of heavy-to-light form factors.  In these calculations
penguin annihilation terms are found to give substantial, imaginary
contributions that correspond to direct CP violation.

The ``charming penguins'' approach \cite{charming}\ incorporates
factorization-violating terms of ${\cal O}(\Lambda_{\mathrm{QCD}}/m_b)$,
especially the penguin terms with charm quarks in the loop.  The small
number of unknown complex amplitudes can be obtained from fits to data.

\section{$B\ra K,\pi$ branching fractions}

\begin{figure}
  \label{fig:Kpi}
  \parbox[b]{.49\linewidth}{
  \noindent\includegraphics[width=0.49\linewidth]{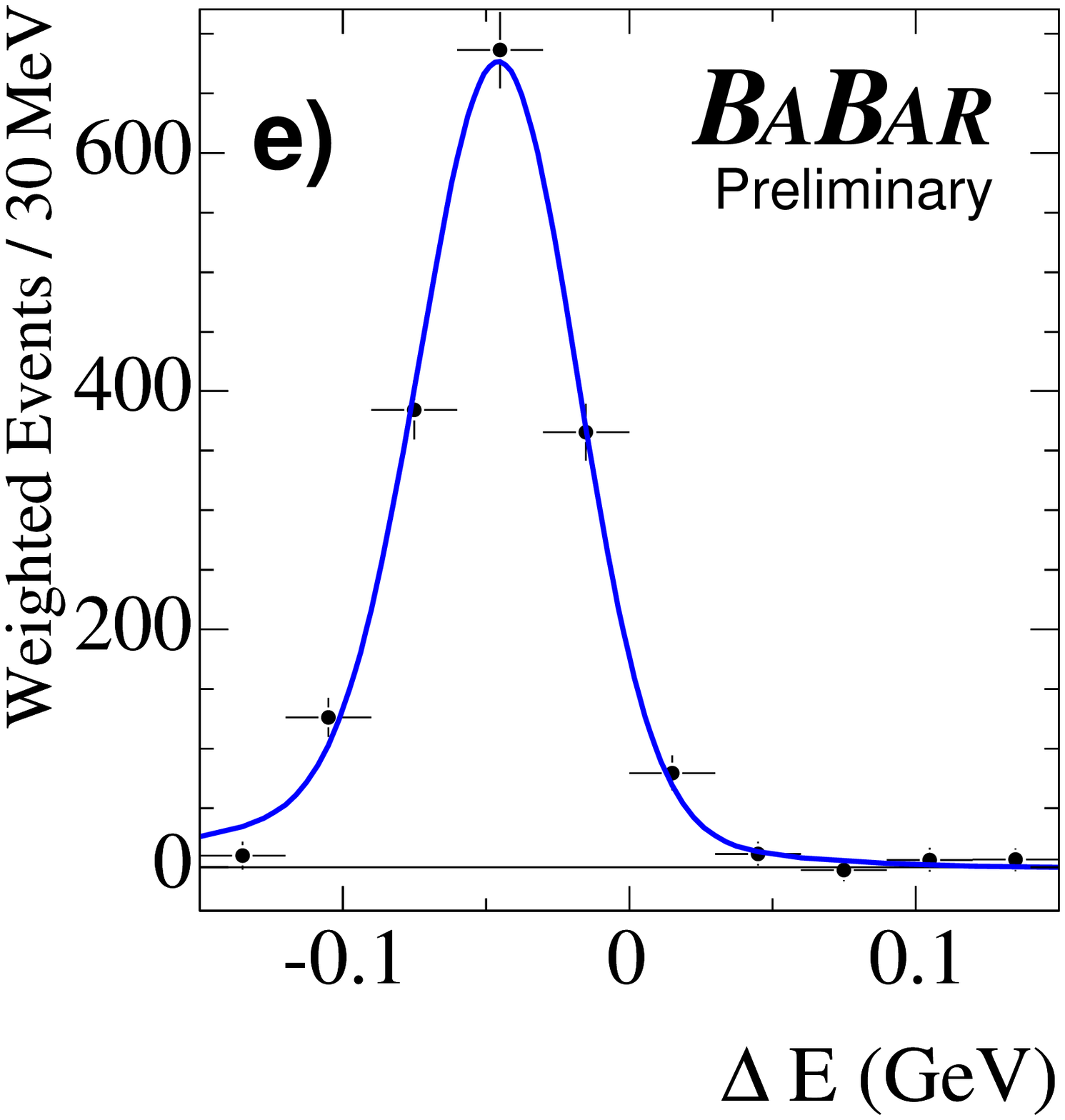}
  \hspace{-5mm}
  \includegraphics[width=0.49\linewidth]{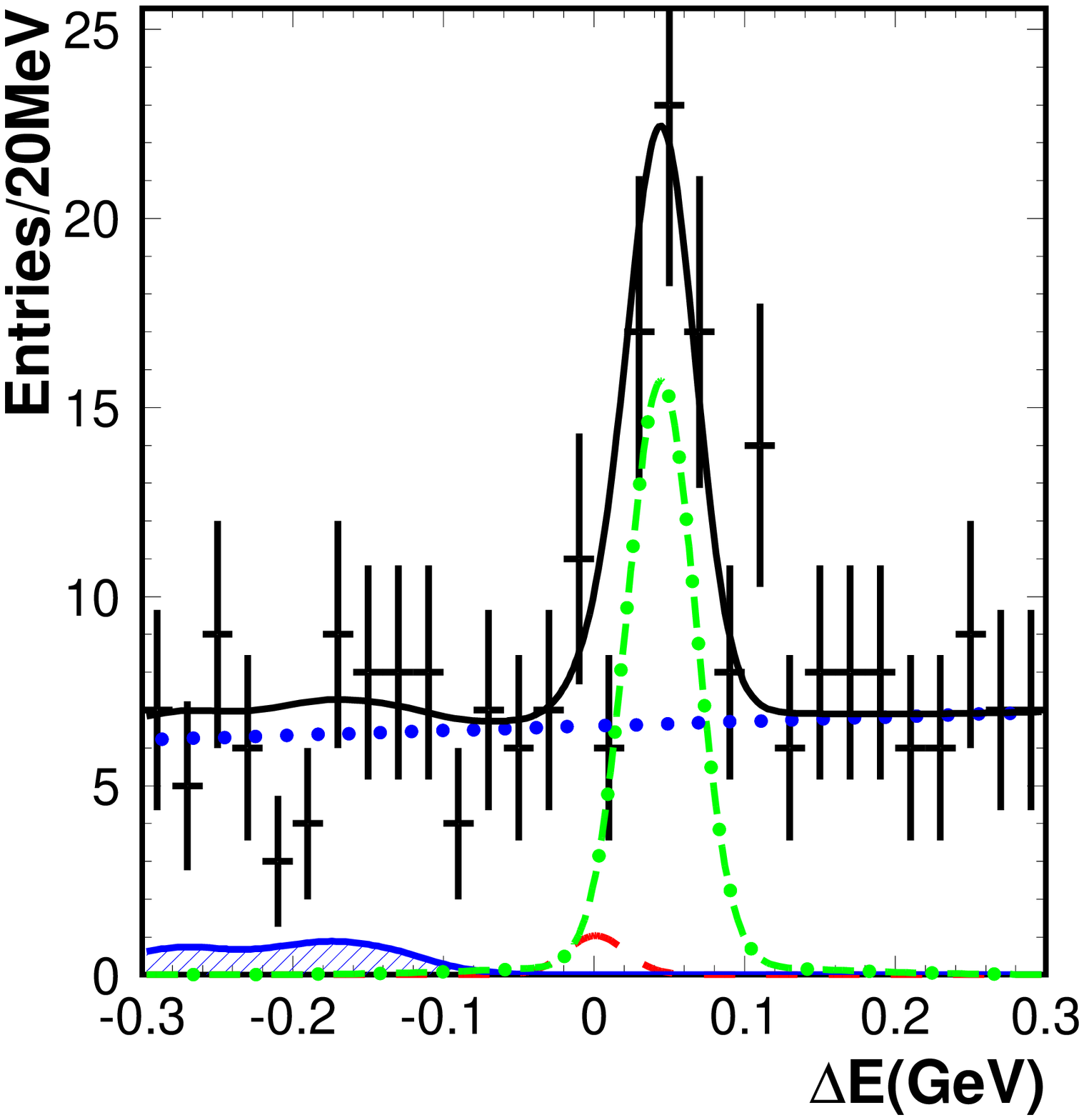}\\
  \includegraphics[width=0.49\linewidth]{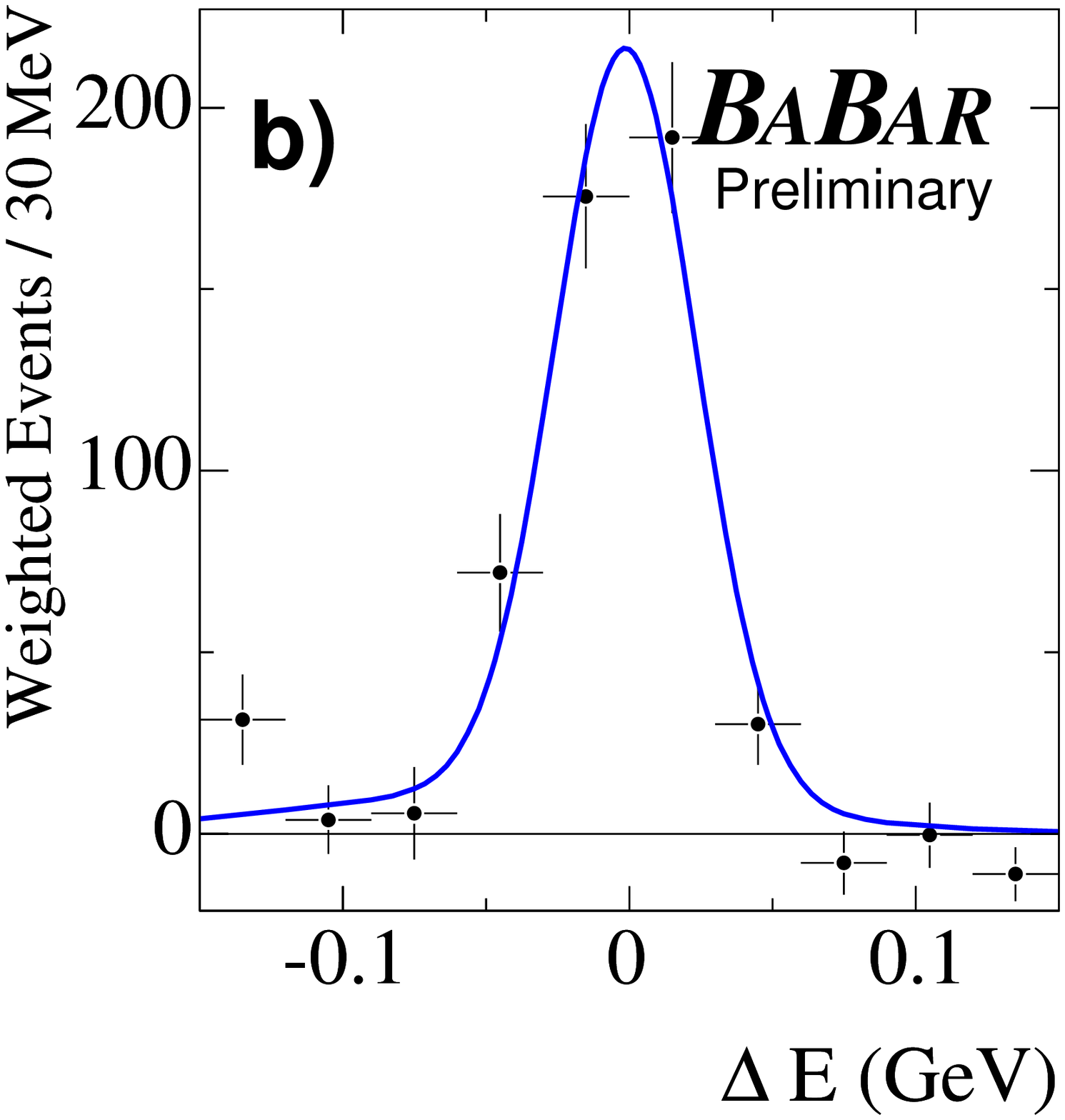}
  \hspace{-5mm}
  \includegraphics[width=0.49\linewidth]{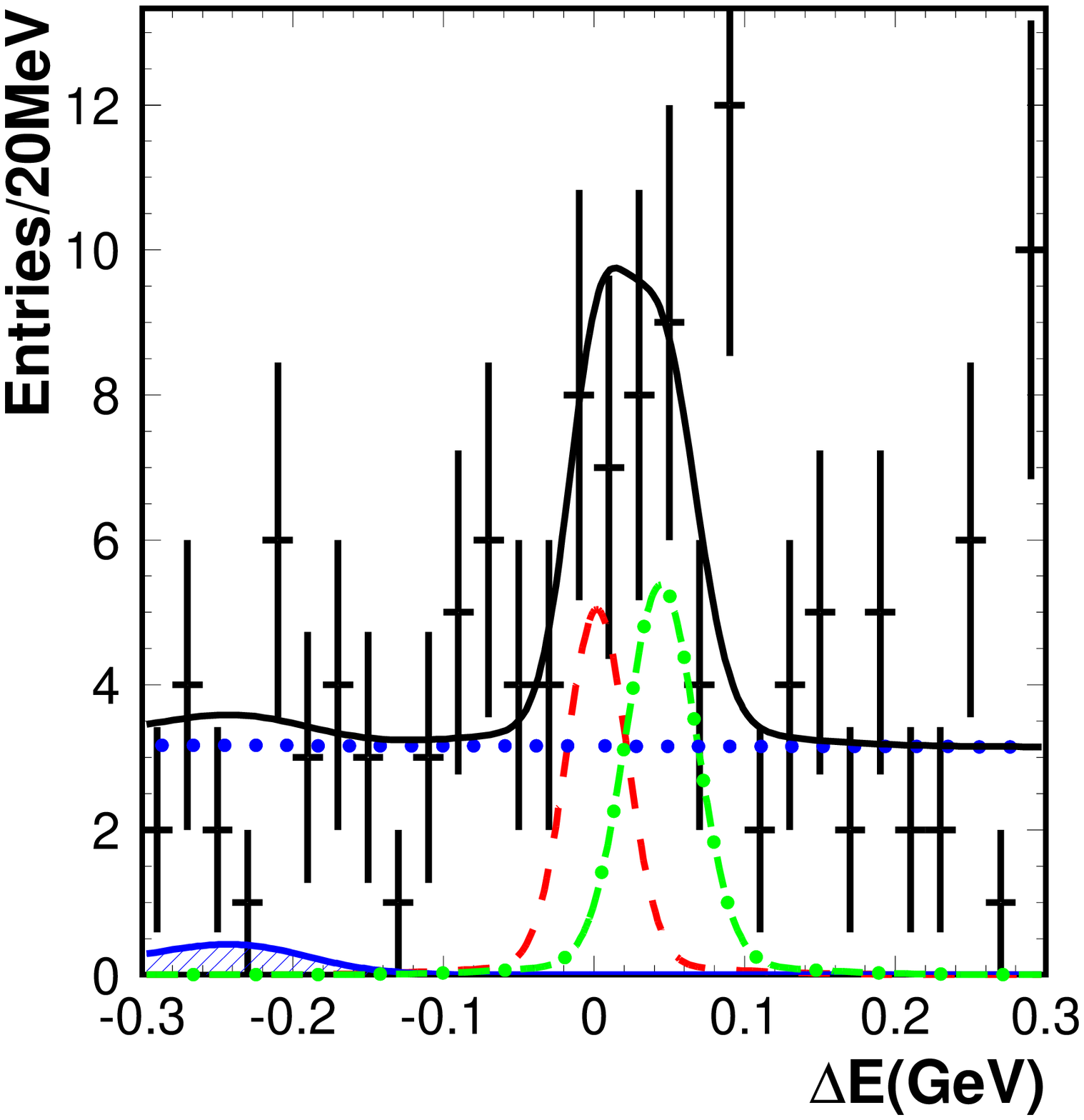}
}%
  \includegraphics[width=0.48\textwidth,bb=0 18 538 526]{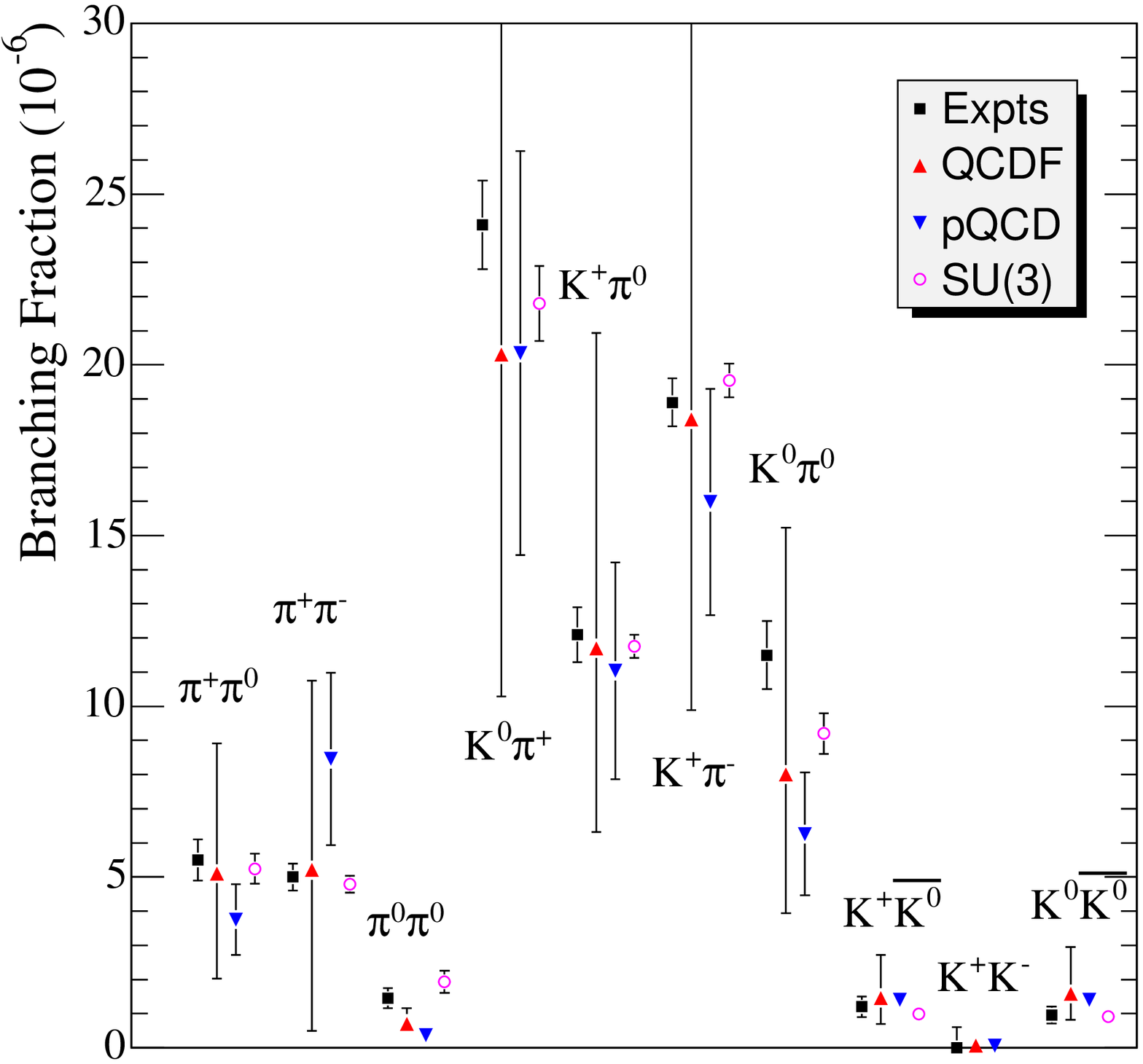}
  \hspace{-15cm}
  \beginpixoverlay[]{425}{200}
  \put(30,180){\colorbox{white}{\phantom{e)}\hspace{-3mm}a)}}
  \put(130,180){c)}
  \put(130,76){d)}
  \put(260,170){e)}
  \end{picture}
  \caption{(a-d) Distributions in \DE\ for two-body $B$
decay candidates.  Signal
sPlots \cite{sPlots} in $\DE_{\pi\pi}$ are given by BaBar \cite{bbHH}\
for (a) 
$\Bz\ra\Kp\pim$ and (b) $\Bz\ra \pip\pim$.  Distributions in
$\DE_{K K}$ are given by Belle \cite{blKK}\ for candidates with
$5.271\ \gev < \mes 
< 5.289\ \gev$ for (c) $\Bz\ra \Kp\Km$ and (d) $\Bp\ra\Kz\pip$.  In (c)
and (d) the dashed red (dot-dashed green) curve is the component from
the fit for the $K K$ signal ($K\pi$ misidentification background).  (e)
World 
average measurements \cite{HFAG} (solid squares) for all $K$, $\pi$ charge
states, with theory estimates from QCDF (``scenario 4'')  \cite{beneke}, triangles; pQCD \cite{keum},
inverted triangles; and flavor-SU(3) \cite{suprunPP}, open squares).}
\end{figure}

The simplest charmless hadronic final states are the ten charge states
among kaons and pions.  These have all been studied with good
sensitivity by BaBar and Belle, who obtain consistent results.  New at
this conference \cite{bbHH} or at LP05 \cite{blKK} are the measurements
given in Table \ref{tab:newKpi}\ and illustrated in
Fig. \ref{fig:Kpi}(a-d).  The decay $\Bz\ra\piz\piz$ has been observed
by both experiments with a branching fraction somewhat larger than
expected (or desired from the point of view of extracting $\alpha_{CKM}$
from an isospin analysis).  Both experiments also have evidence for the
two di-kaon states with a neutral kaon.

\begin{table}[htbp]
 \label{tab:newKpi}
 \begin{tabular}{llcc}
 \hline
     \tablehead{1}{c}{b}{Mode}
   & \tablehead{1}{c}{b}{$\calB\ (10^{-6})$, BaBar}
   & \tablehead{1}{c}{b}{$\calB\ (10^{-6})$, Belle} \\
 \hline
   $\Bz\ra\pip\pim$	& $5.5\pm0.4\pm0.3$  	&    &	\\
   $\Bz\ra\Kp\pim$	& $19.2\pm0.6\pm0.6$ 	&    &	\\
   $\Bz\ra\Kp\Km$	& $<0.40$ (90\% C.L.)	& $<0.37$ (90\% C.L.)	\\
   $\Bp\ra\Kz\Kp$	& 	& $1.0\pm0.4\pm0.1\ (3.0\sigma)$	\\
   $\Bz\ra\Kz\Kzb$	& 	& $0.8\pm0.3\pm0.1\ (3.5\sigma)$	\\
 \hline
 \end{tabular}
 \caption{New $B\ra K,\pi$ branching fraction measurements.}
\end{table} 

The current status of measurements is summarized by the world averages
\cite{HFAG} 
plotted in Fig. \ref{fig:Kpi}(e), alongside phenomenological estimates
from QCDF \cite{beneke}, pQCD \cite{keum}, and flavor SU(3)
\cite{suprunPP}.  We see that on the whole the agreement among the
models and with the data is within the uncertainties.  (In the case of
the flavor-SU(3) global fit, and to a lesser extent for the other
models, the phenomenology is quite strongly influenced by these data.)

The branching fractions span a rather wide range, with the
prominence of penguin amplitudes indicated by the relatively large rate
to the $\Kp\pim$ state.  One puzzling feature of the $K\pi$ family is
the larger than predicted branching fraction measured for $\Kz\piz$.
One might interpret a large $\Kz\piz$ as indicative of an enhanced
electroweak penguin amplitude.  More specifically, the SU(3) prediction
$$
2\frac{{\cal B}(\Bp\ra\Kp\piz) + c.c.}{{\cal B}(\Bp\ra\Kz\pip) + c.c.}
= \frac{1}{2}\frac{{\cal B}(\Bz\ra\Kp\pim) + c.c.}{{\cal
B}(\Bz\ra\Kz\piz) + c.c.}
$$
which recently appeared to be violated by the data, now gives for the
ratio of left to right-hand sides $1.00\pm0.09/0.82\pm0.08 =
1.22\pm0.16$, less than two sigma from unity.  Corrections for
final-state radiation, affecting only modes with charged particles,
reduces the difference between these ratios. 

Experiment and theory agree on the smallness of the $\Kp\Km$ branching
fraction, expected because only weak annihilation (exchange) can account
for it.  In the di-pion system, more theoretical work is needed to
extract useful predictions for $\Bz\ra\piz\piz$.  This mode is mediated
by the $C$ and CKM-suppressed $P$ amplitudes, which may be
constructively interfering.  The calculation in QCDF is sensitive to
unknown input parameters.  A correct prediction was made from the
charming penguin approach \cite{charming}.

\section{Branching fractions for P-P and V-P modes}

\begin{figure}
  \label{fig:dS1}
  \includegraphics[width=0.99\textwidth,bb=0 11 538 256]{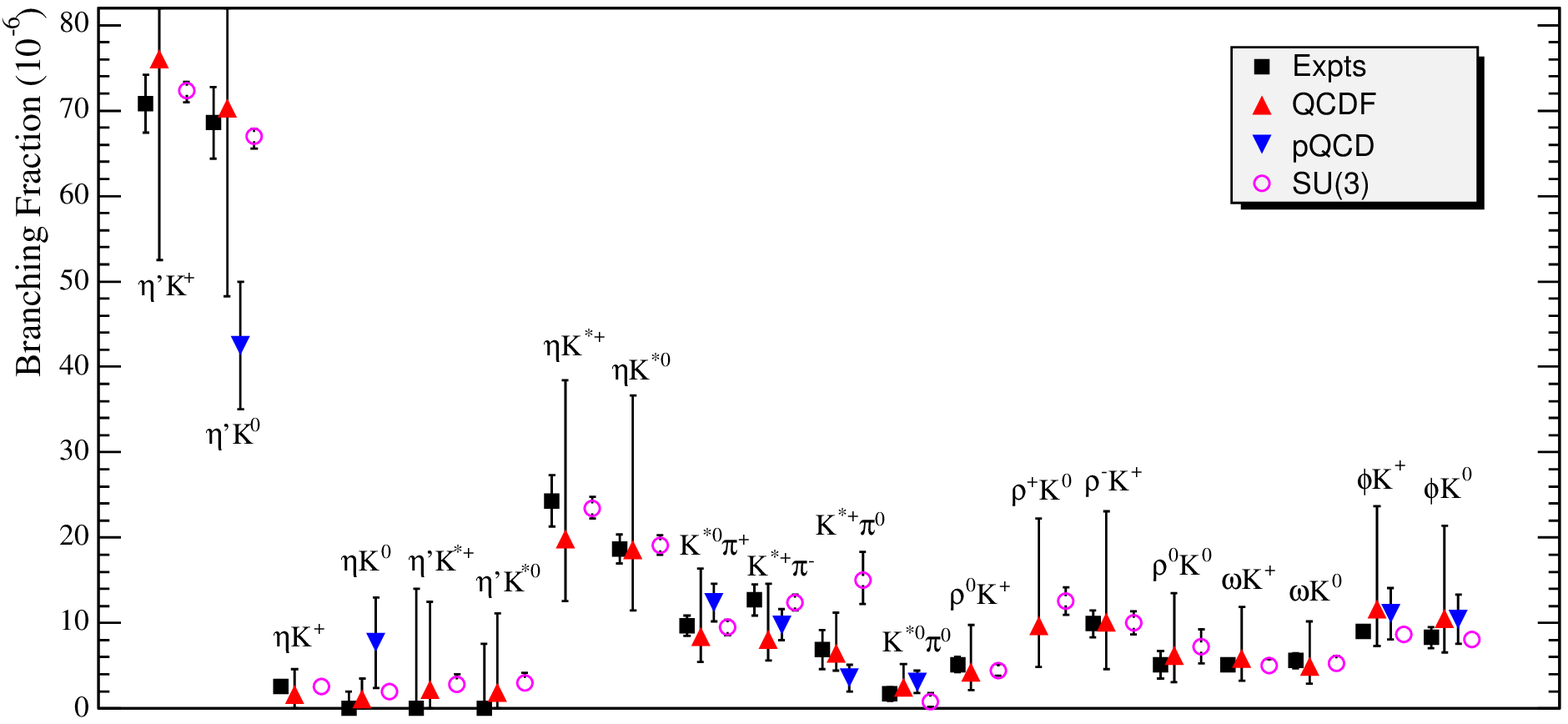}
  \caption{World average measurements \cite{HFAG} (solid squares) for $\Delta S=1$
$P$--$P$ and $V$--$P$ modes, with theory estimates from QCDF (``scenario
4'') \cite{beneke}, triangles;
pQCD \cite{keum,kouSanda}, inverted triangles; and flavor-SU(3)
\cite{suprunPP,suprunVP}, open squares.  The error bar for a point on
the abscissa indicates the 90\% C.L. upper limit.} 
\end{figure}

Investigations, both experimental and theoretical, have been made for
most two-body combinations of particles belonging to the ground-state
pseudoscalar and vector nonets.  We show in Fig.\ \ref{fig:dS1}\ the
world-average measurements of branching fractions for the $|\Delta S| = 1$
$P$--$P$ and $V$--$P$ modes (excluding $(K,\pi)$ covered above),
together with some of 
the phenomenological estimates as in the previous section.  These modes
tend to be dominated by $b\ra s$ penguin amplitudes, since trees are
Cabibbo suppressed.

The largest charmless two-body hadronic branching fractions are measured
for the \etapK\ states.  There has been much discussion in the
literature of the mechanisms possibly responsible for enhancing these
modes.  The \etapr\ is close to being a flavor-singlet state, allowing
its coupling to pure glue, so the direct hard-glue amplitude $S^\prime$
may play a role, as is found in the global SU(3)-based fits.  However,
other states with \etapr\ such as \etapKst\ are not correspondingly
large, as one sees in Fig.\ \ref{fig:dS1}.  Instead, the \etaKst\ mode
is quite strong, and again, \etaK\ weak.  The relative strengths of
these four modes supports an early conjecture of Lipkin \cite{lipkin}:
coherent combination of the penguins $b\ra s u\ubar$ and $b\ra s
s\sbar$, strongly constructive for \etapK, destructive for \etaK.  The
relative strength for $\eta$ and $\etapr$ recoiling against \Kst\ is
reversed, because of the odd orbital parity in $V$--$P$ final states.
The QCDF calculation reproduces the data rather well, when the optimal
choices for parameters such as the strange quark mass and form factors
are taken.  The pQCD prediction range for \etapKz\ is somewhat low.  The
SU(3) fit has enough freedom to match the data well.  The remaining
$|\Delta S|=1$ mode measurements and predictions are in fairly good
agreement among the theoretical estimates and with experiment.

\begin{figure}
  \label{fig:dS0}
  \includegraphics[width=0.99\textwidth,bb=0 11 538 256]{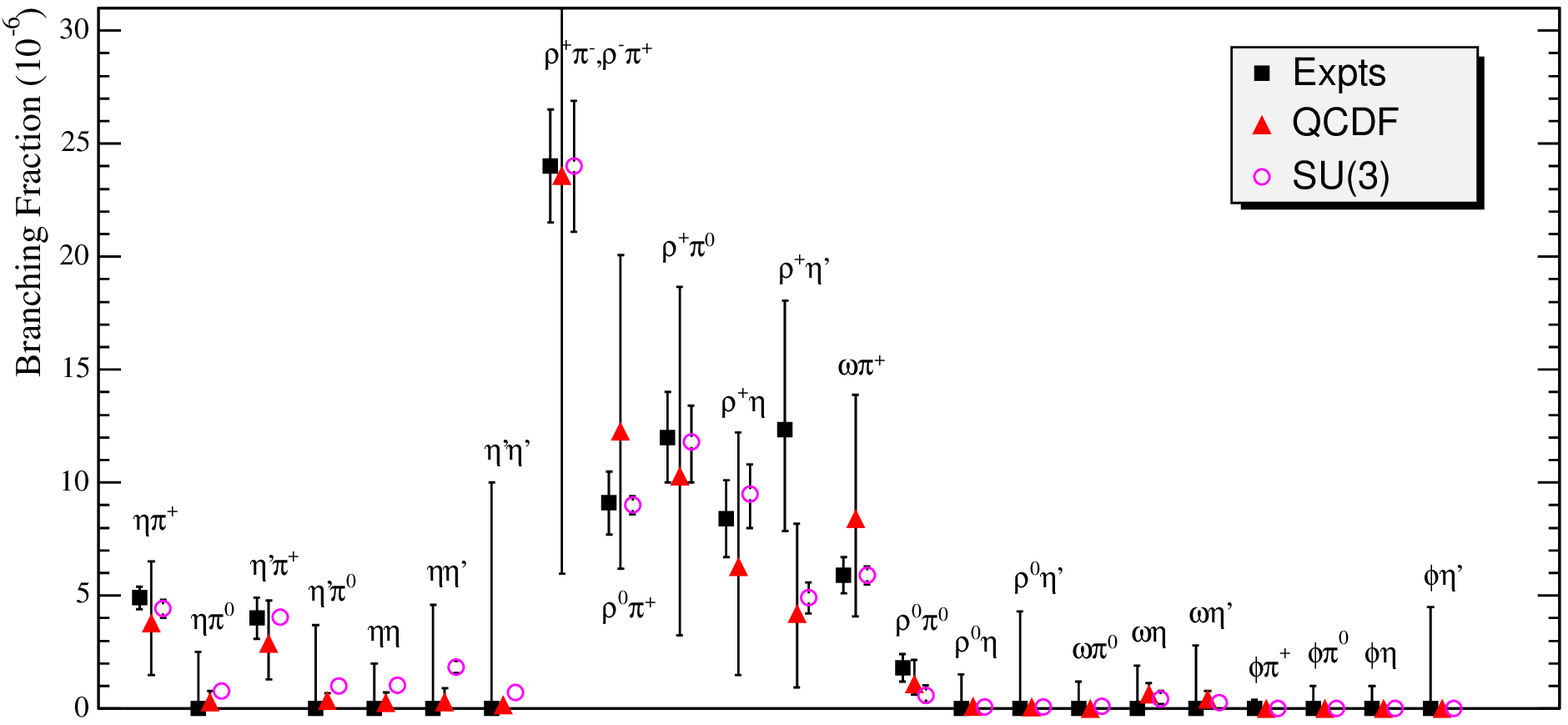}
  \caption{World average measurements \cite{HFAG} (solid squares) for $\Delta S=0$
$P$--$P$ and $V$--$P$ modes, with theory estimates from QCDF (``scenario
4'') \cite{beneke}, triangles;
and flavor-SU(3)
\cite{suprunPP,suprunVP}, open squares.  The error bar for a point on
the abscissa indicates the 90\% C.L. upper limit.} 
\end{figure}

The corresponding display of measurements compared with theory for the
$\Delta S=0$ modes is given in Fig. \ \ref{fig:dS0}.  The tree-dominated
$\rho\pi$ and $\omega\pi$ states stand out, along with \fetarhop,
(possibly) \fetaprhop, and $\eta^{(\prime)}\pip$.  The decays with
all-neutral final state particles must proceed through color-suppressed
tree, $b\ra d$ penguin, or flavor-singlet penguin amplitudes, and of
these only $\rhoz\piz$ has been observed.  Experimental values and
limits on the various combinations of $\eta,\ \etapr$, and \piz\ bear on
the interpretation of time-dependent $CP$ measurements in $\etapKz$.
Specifically, these limits imply bounds on the ``tree pollution''
correction $\Delta S =
\sin{2\beta_\mathrm{eff}}-\stwob$, enabling precision tests of the
consistency between $b\ra\s$ penguin and $b\ra c\cbar s$ determinations
of \stwob.

\section{Three-body decays}

\begin{figure}
  \label{fig:Kpipi}
  \includegraphics[width=0.48\textwidth,bb=10 13 515 356]{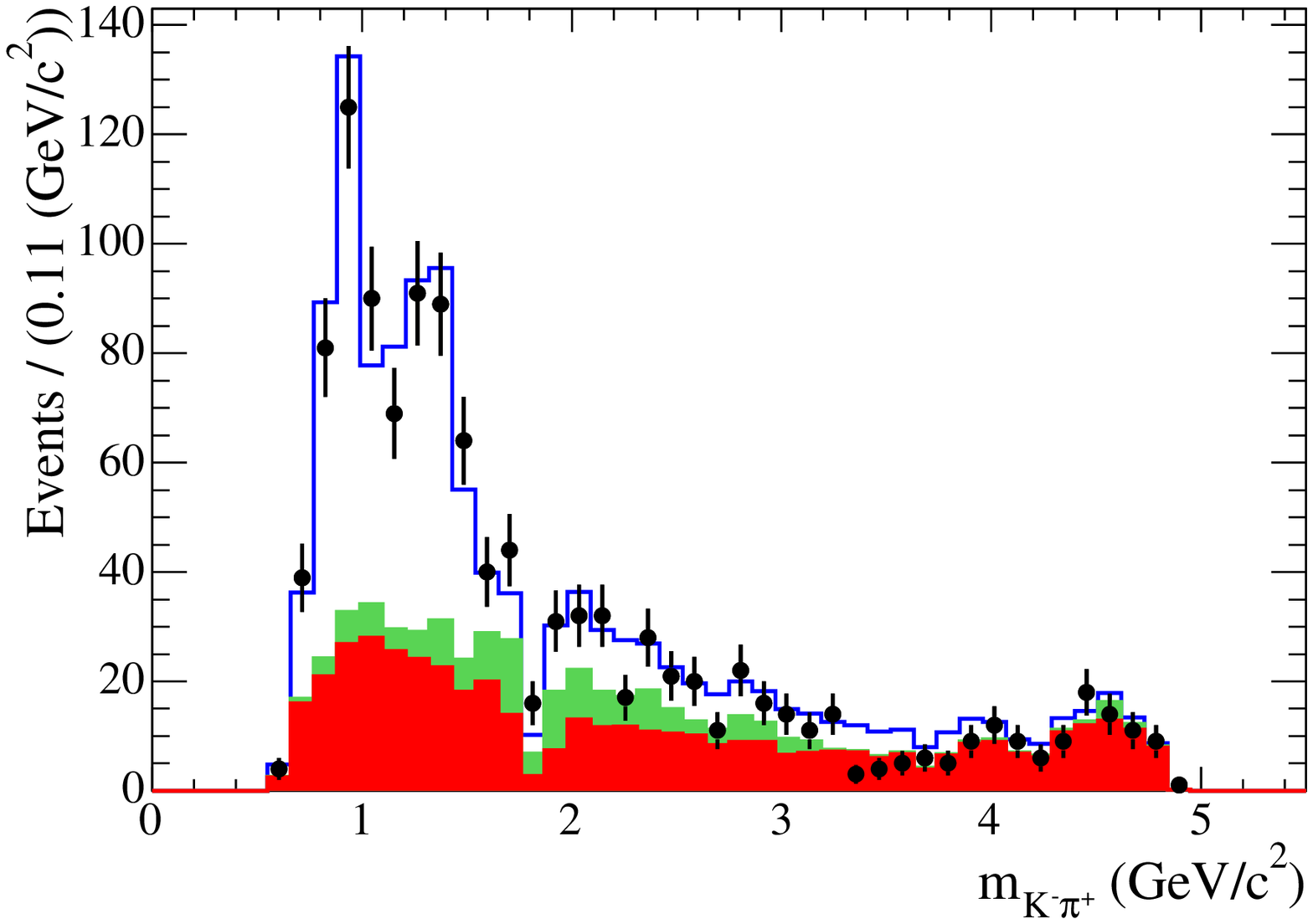}\quad
  \includegraphics[width=0.48\textwidth,bb=10 13 515 356]{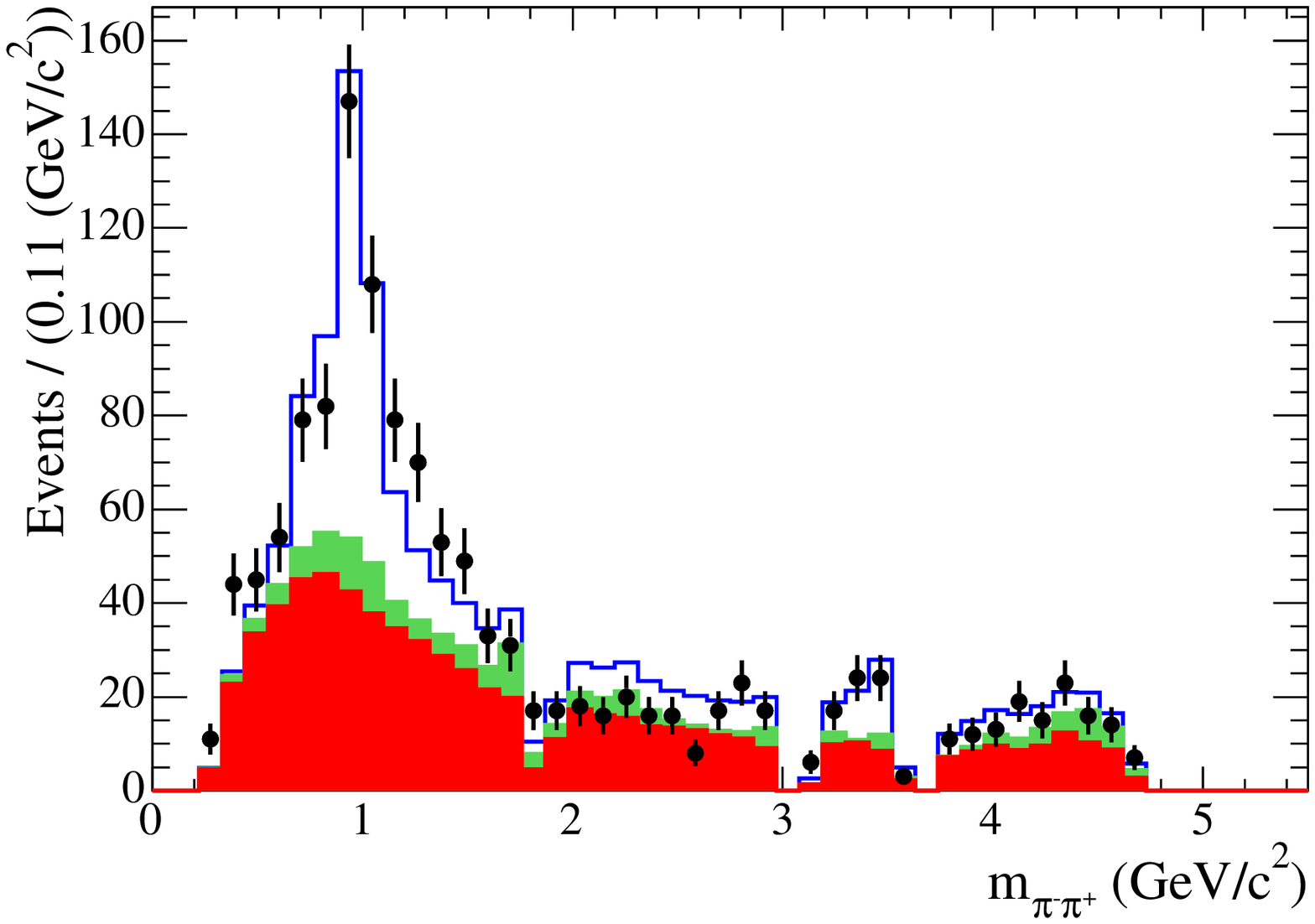}
  \caption{Dalitz plot projections on $m(\Km\pip)$ and
$m(\pim\pip)$ for $\Bm\ra\Km\pip\pim$ data from BaBar.  For the
$m(\Km\pip)$ or $m(\pim\pip)$ distribution the other mass combination
($m(\pim\pip)$ or $m(\Km\pip)$) is required to be greater than $2\ \gev$.} 
\end{figure}

A number of three-body charmless $B$ decays have been
observed: $\Bp\ra\phi\phi\Kp$, with a branching fraction of
$2.5\times10^{-6}$, the various charge states of three kaons (6 to
$30\times10^{-6}$) and of $K\pi\pi$ (35 to $55\times10^{-6}$), and
$\pip\pip\pim$ ($16\times10^{-6}$) \cite{HFAG}.
Generally rather large continuum backgrounds must be dealt with for
these inclusive final states.  Nonetheless, for several of them detailed
analyses of the Dalitz distributions have been performed.  An example is
the final state $\Kpm\pipm\pimp$, published earlier this year by Belle
\cite{blKpipi} on 152 million \BB\ pairs, and giving
$\calB(\Bpm\ra\Kpm\pipm\pimp)=(46.6\pm2.1\pm4.3)\times10^{-6}$, and
presented at LP05 by BaBar \cite{bbKpipi} (226 million \BB\ pairs) and
giving $\calB(\Bpm\ra\Kpm\pipm\pimp)=(64.1\pm2.4\pm4.0)\times10^{-6}$.
Mass projections from BaBar's analysis for the \Bm\ decay are given in
Fig.\ \ref{fig:Kpipi}.  Both experiments find significant signals within
this final state for $\Kstarz(890)\pipm$, $S$-wave $\Kstarz(1430)\pipm$,
$\rhoz(770)\Kpm$, and $f_0(980)\Kpm$.  A similar analysis of
$\Bpm\ra\pipm\pip\pim$ has been performed recently by BaBar \cite{pipipi}.

\section{Direct CP-violation charge asymmetry measurements}

The appearance of direct \CP-violating effects, those coming from the
decay amplitude alone, requires a difference of both weak and strong
phases between two terms, as may occur for example if both tree and
penguin contributions are present.  The weak phases change sign relative
to the strong phases under the \CP\ operation.  For reasonable
sensitivity these weak and strong phase differences also need to be of
comparable magnitude.  Typically the greater the direct \CP\ sensitivity
of a given decay mode, the smaller its rate, making the experimental
observation difficult.  So far the observed instances occur in Belle's
measurement of the $C$ term in the time-dependence of $\Bz\ra\pip\pim$
discussed in Sakai's talk, and in the charge asymmetry \acp\ in
$\Bz\ra\Kpm\pimp$.  In the past year both BaBar and Belle have reported
non-vanishing charge asymmetries in this decay.  The BaBar measurement
gives $\acp = -0.133\pm0.030\pm0.009$ in about 1600 decays from
$230\times10^6$ \BB\ pairs \cite{bbKpiAcp}; an update of Belle's
measurement to a sample from $386\times10^6$ \BB\ pairs \cite{blKpiAcp}
presented at LP05 gives $\acp = -0.113\pm0.022\pm0.008$.  The world
average \cite{HFAG}, which includes also measurements by CLEO and CDF,
is given in the first line of Table \ref{tab:Acp}, along with
predictions from the several theoretical approaches.

The table gives results for several related modes, as well
as other charmless modes for which the measurement precision is reaching
an interesting sensitivity, or there is an expectation that interference
between two weak amplitudes of comparable strength may produce a
measurable effect, as in $\eta\Kpm$ and $\eta\pipm$.  The apparent
difference in \acp\ between the $\Bz\ra\Kpm\piz$ and $\Bz\ra\Kpm\pimp$
modes is not understood at present.

\begin{table}[htbp]
 \label{tab:Acp}
 \begin{tabular}{lcccc}
 \hline
     \tablehead{1}{c}{b}{Mode}
 & \tablehead{1}{c}{b}{\acp, exptl. world ave.}
 & \tablehead{1}{c}{b}{\acp, QCDF}
 & \tablehead{1}{c}{b}{\acp, pQCD}
 & \tablehead{1}{c}{b}{\acp, $SU(3)$} \\
 \hline
 $\Bz\ra K^\pm\pi^\mp$	& $-0.115\pm0.018$ 	& $-0.04\pm0.01$   & $-0.129\sim-0.219$	& $-0.096^{+0.017}_{-0.011}$	\\
 $\Bz\ra K^\pm\piz$	& $+0.04\pm0.04$ 	& $-0.04\pm0.01$   & $-0.100\sim-0.173$	& $+0.02\pm0.02$	\\
 $\Bz\ra \Kz\pi^\pm$	& $-0.02\pm0.04$ 	& $+0.003\pm0.007$ & $-0.006\sim-0.015$	& $-0.00\pm0.01$	\\
 $\Bpm\ra \eta\pi^\pm$	& $-0.11\pm0.08$ 	& $+0.056\pm0.20$  &	& $-0.05^{+0.06}_{-0.05}$	\\
 $\Bpm\ra \eta K^\pm$	& $-0.33\pm0.12$ 	& $+0.10\pm0.30$   &	& $-0.39\pm0.04$	\\
 $\Bpm\ra \etapr K^\pm$	& $+0.031\pm0.021$ 	& $-0.008\pm0.040$ &	& $+0.003^{+0.005}_{-0.003}$	\\
 $\Bpm\ra \eta \Kstarpm$	& $+0.03\pm0.11$ 	& $-0.057\pm0.21$  &	& $0.00\pm0.02$	\\
 $\Bz\ra \eta \Kstz$	& $-0.01\pm0.08$ 	& $+0.008\pm0.04$  &	& $+0.01\pm0.006$	\\
 $\Bpm\ra\pip\pim\pipm$	& $+0.01\pm0.09$ 	&    &	&	\\
 $\Bpm\ra\rhoz\pipm$	& $-0.07^{+0.12}_{-0.13}$ 	& $-0.11\pm0.19$	&	& $-0.16\pm0.04$	\\
 $\Bpm\ra\pip\pim\Kpm$	& $+0.022\pm0.029$ 	&    &	&	\\
 $\Bpm\ra\rhoz\Kpm$	& $+0.31^{+0.12}_{-0.11}$ 	& $+0.32\pm0.60$    &	& $+0.21\pm0.10$	\\
 $\Bpm\ra f_0(980)\Kpm$	& $-0.020^{+0.068}_{-0.065}$ 	&    &	&	\\
 $\Bpm\ra\Kstarz\pipm$	& $-0.093\pm0.060$ 	& $+0.008\pm0.003$    &	& 0	\\
 \hline
 \end{tabular}
 \caption{Direct $CP$-violating charge asymmetry measurements and
theoretical estimates.}
\end{table} 

\section{Summary and conclusions}

In this brief review I have omitted discussion of many developments of
interest in the area of rare $B$ decays.  Important work has been done
with vector-vector decays, where there is an unresolved puzzle with the
failure of transverse amplitude suppression in penguin-dominated modes
such as $\Bz\ra\phi\Kstarz$.  The study of exclusive and inclusive
radiative charmless decays contributes to the determination of CKM
matrix elements (see Kinoshita's talk), provides tests of the theory and
limits on new physics.  Some final states with charmless baryons have
been observed.

My emphasis here has been on the charmless mesonic decays.  With a
modest number of free parameters the calculations now available are able
to describe quite well the large and growing collection of measured
branching fractions and \CP\ asymmetries.  Uncertainties in the models
are still rather large, and need to be improved to facilitate
unambiguous tests of the Standard Model with the full statistical power
of near future measurements.


\begin{theacknowledgments}
It is a pleasure to thank Jim Smith, Jon Rosner, and Yoshi Sakai for
helpful discussions, and Jim Olsen and Owen Long on whose compilations I
have drawn.  This work is supported in part by the Department of Energy
under grant DE-FG02-04ER41290.
\end{theacknowledgments}


\bibliographystyle{aipprocl} 

\begin{thebibliography}{99}

\bibitem{blBtaunu}
Belle Collaboration, K. Abe et al., BELLE-CONF-0566 (2005) [hep-ex/0507034].

\bibitem{bbBtaunu}
BaBar Collaboration, B. Aubert et al., BABAR-CONF-05/034 (2005)
[hep-ex/0507069]. 

\bibitem{LQCDfB}
S. M. Ryan, \npps{106}, 86 (2002).

\bibitem{bib:ckmFitBtaunu}
CKMfitter Group (J. Charles et al.),
Eur. Phys. J. C41, 1-131 (2005) [hep-ph/0406184],
updated results and plots available at: \url{http://ckmfitter.in2p3.fr}.

\bibitem{suprunPP}
C-W. Chiang et al., \jprd{70}, 034020 (2004) [hep-ph/0404073];
J. Rosner, private communication.

\bibitem{suprunVP}
C-W. Chiang et al., \jprd{69}, 034001 (2004) [hep-ph/0307395].

\bibitem{bblRMP}
M.~Bauer, B. Stech, and M. Wirbel, \zpc{34}, 103 (1987); 
G. Buchalla, A. Buras, and M. E. Lautenbacher, \jrmp{68},
1125 (1996). 

\bibitem{ali}
A.~Ali and C.~Greub, \jprd{57}, 2996 (1998);
A.~Ali, G.~Kramer, and C.~D.~L\"u, \jprd{58}, 094009  (1998).

\bibitem{BBNS}
M. Beneke, G. Buchalla, M. Neubert, and C. T. Sachrajda, \jprl{83}, 1914
(1999) [hep-ph/9905312].

\bibitem{beneke}
M. Beneke and M. Neubert, \npb{675}, 333 (2003) [hep-ph/0308039].

\bibitem{pQCD}
G.~P.~Lepage and S.~Brodsky, \jprd{22}, 2157 (1980); 
J.~Botts and G.~Sterman, \npb{325}, 62 (1989);
Y.~Y.~Keum \etal, \plb{504}, 6  (2001), \jprd{63}, 054006  (2001);
Y.~Y.~Keum and H.~N.~Li, \jprd{63}, 074008 (2001).

\bibitem{keum}
Y.-Y. Keum., Pramana 63, 1151-1170 (2004) [hep-ph/0410337].

\bibitem{kouSanda}
E. Kou and A. I. Sanda, \plb{525}, 240 (2002).

\bibitem{charming}
M. Ciuchini et al., \npb{501}, 271 (1997) [hep-ph/9703353]; \plb{515},
33 (2001) [hep-ph/0104126].

\bibitem{sPlots}
M. Pivk and F. R. Le Diberder, \nima{}\ (2004, in press) [physics/0402083].

\bibitem{bbHH}
BaBar Collaboration, B. Aubert et al., BABAR-CONF-05/013
(2005) [hep-ex/0508046].

\bibitem{blKK}
Belle Collaboration, K. Abe et al., BELLE-CONF-0524 
(2005) [hep-ex/0506080]. 

\bibitem{HFAG}
Heavy Flavor Averaging Group,
\url{http://www.slac.stanford.edu/xorg/hfag/}; data from the ``LepPho
2005'' tables. 

\bibitem{lipkin}
H.\ J.\ Lipkin, \plb{254}, 247 (1991).

\bibitem{blKpipi}
Belle Collaboration, A. Garmash et al., \jprd{71}, 092003 (2005).

\bibitem{bbKpipi}
BaBar Collaboration, B. Aubert et al., \jprd{}(in press),
BABAR-PUB/05/027 [hep-ex/0507004].

\bibitem{pipipi}
BaBar Collaboration, B. Aubert et al., \jprd{72} 052002 (2005).

\bibitem{bbKpiAcp}
BaBar Collaboration, B. Aubert et al., \jprl{93}, 131801 (2004).

\bibitem{blKpiAcp}
Talk by K. Abe at LP2005; \jprl{93}, 191802 (2004).

\end{thebibliography}

\end{document}